\documentclass[english]{article}
\usepackage{jcappub}
\usepackage{lmodern}

\usepackage{slantsc}
\usepackage{graphicx}
\DeclareGraphicsExtensions{.pdf,.png,.jpg}
\graphicspath{{./}}

\def \obsT{\widetilde{\Delta T}}
\def \obsC{\widetilde{C}}
\def \unbC{\bar{C}}
\def \unbsC{\bar{C}^\text{scalar}}
\def \trueC{C^\text{true}}
\def \fidC{C^\text{fid}}
\def \obsa{\widetilde{a}}
\def \q{\hat{\mathbf q}}
\def \d{\mathrm{d}}
\def \lmax{l_\text{max}}
\def \cov{\text{Cov}}

\def \Planck{\textsl{\textsc{Planck}}}

\begin{document}

\title{Effect of noncircularity of experimental beam on CMB parameter estimation}

\author[a]{Santanu Das,}
\author[a]{Sanjit Mitra}{
\author[b]{and Sonu Tabitha Paulson}

\emailAdd{santanud@iucaa.ernet.in}
\emailAdd{sanjit@iucaa.ernet.in}
\emailAdd{sonu.tabitha@gmail.com}

\affiliation[a]{IUCAA, P. O. Bag 4, Ganeshkhind, Pune 411007, India}
\affiliation[b]{University of Madras, Chennai-25, India}

\keywords{CMB analysis}

\abstract{
Measurement of Cosmic Microwave Background (CMB) anisotropies has been playing a lead role in precision cosmology by providing some of the tightest constrains on cosmological models and parameters. However, precision can only be meaningful when all major systematic effects are taken into account. Non-circular beams in CMB experiments can cause large systematic deviation in the angular power spectrum, not only by modifying the measurement at a given multipole, but also introducing coupling between different multipoles through a deterministic bias matrix. Here we add a mechanism for emulating the effect of a full bias matrix to the {\Planck} likelihood code through the parameter estimation code SCoPE. We show that if the angular power spectrum was measured with a non-circular beam, the assumption of circular Gaussian beam or considering only the diagonal part of the bias matrix can lead to huge error in parameter estimation. We demonstrate that, at least for elliptical Gaussian beams, use of scalar beam window functions obtained via Monte Carlo simulations starting from a fiducial spectrum, as implemented in {\Planck} analyses for example, leads to {\em only few percent of sigma deviation} of the best-fit parameters. However, we notice more significant differences in the posterior distributions for some of the parameters, which would in turn lead to incorrect errorbars. These differences can be reduced, so that the errorbars match within few percent, by adding an iterative reanalysis step, where the beam window function would be recomputed using the best-fit spectrum estimated in the first step.
}

\maketitle

\section{Introduction}

Measurements of Cosmic Microwave Background (CMB) opened a whole new era in theoretical physics. Not only it made the standard cosmological model widely acceptable, but it offered precise measurements of cosmological parameters. Several high precision ground based and space based experiments were carried out in the past decades to measure CMB anisotropies. Recently, WMAP ~\cite{Hinshaw2007,wmap7-anomalies,WMAP9-cosmology,WMAP9-anomalies} and {\Planck}~\cite{planck16,planck23} mapped the full CMB sky with few arcmin level resolutions.
However, high precision measurements demand accurate accounting of the systematic errors. Systematic errors can arise at different stages in CMB analysis, such as, foreground cleaning~\cite{Tegmark2003,PlanckCompSep-13}, instrument calibration~\cite{LFICalibration-13,HFICalibration-13}, measurement of beam response~\cite{Huffenberger10,MCBR} and estimation of the angular power spectra~\cite{PlanckLike-13} from the observed maps. In this paper we focus on the effect of noncircular beams~\cite{Souradeep2001,Fosalba2002,Mitra2004,Souradeep2006,Mitra2009,febecop,LFIBeams-13,HFIBeams-13}, one of the most important and challenging sources of systematic error.

Observed CMB sky is a convolution of the underlying true sky with the instrumental beam response function. Accounting for the beam is thus necessary for measuring the statistical properties of the true sky. This turns out to be trivial if the beam is symmetric about the pointing direction, an unbiased estimator of the true angular power spectrum can be readily obtained.
However, an actual beam response function is never so symmetric.  The intrinsic optics of the instruments, aberrations due to the placement of detectors away from the principle optic axis, non-uniform distribution of pointing directions in a pixel, finite sampling duration in scanning and many other effects can distort a beam to make it non-circular.  The asymmetries become progressively important in different analyses as the experiments strive to extract almost all the information embedded in the anisotropies.
%
Beam asymmetry may lead to spurious effects in the measured CMB data. It can bias the angular power spectrum in a non-trivial way by introducing coupling between different multipoles through a bias matrix~\cite{Mitra2004}. It can also cause statistical isotropy violation in the CMB anisotropy maps~\cite{Hanson2010,Das2013,Das2014a,Joshi2012}. Here we study how the distortions in the angular power spectra caused by non-circular beams affect cosmological parameter estimation.

If the bias matrix that couples different multipoles was available, it appears obvious that one would use it to obtain an unbiased estimator of angular power spectra and perform parameter estimation starting from the unbiased estimator. However, this process would encounter two major hurdles. First, attempt to unbias the estimator will introduce covariances among different multipoles. Second, perhaps the more important one, it is highly non-trivial to obtain the bias matrix for an experiment with non-circular beam shapes and complex scanning strategy. In literature the near-diagonal components of the bias matrix has been computed to leading order for simple scanning strategies only for temperature anisotropies~\cite{Mitra2004,Mitra2009}, though efforts are on to extend the method to polarization also~\cite{Fidy}.
Our present work thus, in fact, evaluates the need for investing (enormous) effort in getting the full bias matrix for experiments like {\Planck} in order to perform unbiased analyses.

Here we develop a mechanism to include the effect of a bias matrix in the parameter estimation code SCoPE \cite{Das2014} in conjunction with the {\Planck} likelihood code~\cite{PlanckLike-13,PlanckES-13}. We use {\Planck} likelihood code in order to get realistic representation of noise. However, instead of deconvolving the observed power spectrum to get an unbiased estimator, we apply ``inverse distortion'' to the theoretical spectra in the likelihood code to get the desired posterior distributions. This step alleviates the need for incorporating a non-trivial covariance matrix. We use different scalar transfer functions to do parameter estimation from the same convolved spectra and study the differences in posterior distributions with the correct estimates.

It is very common in CMB analysis to use an effective scalar window function, which inherently assumes the beam to be azimuthally symmetric about the pointing direction. However, if the effective window function was derived using Monte Carlo simulations \cite{febecop}, including the full details of beams and scanning strategy, as was done for {\Planck} analyses, one may be able to deceive an analysis by closely emulating the effect of a non-circular beam, as long as the fiducial power spectrum used for the simulations is close to the true one. Here we use WMAP and Planck best-fit power spectra to get two different scalar transfer functions and use them for estimating parameters from a test observed power spectrum obtained by convolving the {\Planck} best fit spectrum with a bias matrix.
However, since a full bias matrix for a mission like {\Planck} does not exist, for numerical computation we must limit ourselves to a case where it is available. We use an elliptical Gaussian beam of similar size and ellipticity as one of the {\Planck} high frequency detectors and non-rotating scan pattern.


This paper is organised as follows. We present a brief primer on the connection between non-circular beams and window function in section~\ref{sec:primer}. The general strategy for studying the effect of non-circular beam on cosmological parameters and the results of our study are included in section~\ref{sec:results}. Section~\ref{sec:conclusion} presents the conclusion and discussions.

\section{Effect of instrumental beams on CMB angular power spectrum}
\label{sec:primer}


Different experiments observe the CMB anisotropy field by scanning the sky through an instrumental beam of finite resolution. The observed time ordered data (TOD) is passed through a refined pixel binning procedure which is generally kept unaware of the beam shapes. This is because deconvolution of all the TOD samples (few trillions for {\Planck} detectors) is computationally prohibitive. The resultant map can then be expressed as a convolution of the true sky with an effective beam function. If the effective beam for every direction was the same and symmetric about each direction, one could show that the angular power spectrum of the measured map $\widetilde{C}_l$ is trivially biased, $\obsC_l = B_l^2 \, C_l$, where $C_l$ is the angular power spectrum of the true sky and $B_l$ is the Legendre transform of the azimuthally symmetric beam function. It is indeed very common in CMB analysis to use an effective scalar window function $B_l^2$. However, perhaps for every CMB experiments, beams are non-circular. Non-circular beams make the effective window function tensorial, $\obsC_l = \sum_{l'} A_{ll'} \, C_{l'}$, as we briefly review below.

\subsection{Observed angular power spectrum}

The observed CMB anisotropy $\obsT(\q)$ in a given direction $\q$ can be expressed as the convolution of the true sky $\Delta T (\q)$ with the instrumental beam function $B(\q,\q')$ plus additive noise $n(\q)$,
\begin{equation}
\obsT(\q) \ = \ \int_{\rm S^2} \d \Omega_{\q'} \, B(\q,\q') \, \Delta T (\q') \ + \ n(\q) \, ,
\end{equation}
where $\d \Omega_{\q}$ is an infinitesimal solid angle around the direction unit vector $\q$.
Using the Spherical Harmonic transform of the observed map
\begin{equation}
\obsa_{lm} \ := \ \int_{\rm S^2} \d \Omega_{\q} \, Y_{lm}(\q) \, \obsT (\q) \ = \  \int_{\rm S^2} \d \Omega_{\q} \int_{\rm S^2} \d \Omega_{\q'} Y_{lm}(\q) [ B(\q,\q') \, \Delta T (\q') \, + \, n(\q) ] \, ,
\end{equation}
and assuming Statistical Isotropy (SI) of the true sky, it can be shown~\cite{Mitra2004} that the expected power spectrum of the observed map can be expressed as
\begin{eqnarray}
\langle \obsC_l \rangle &=& \frac{1}{2 l + 1} \sum_{m=-l}^l \langle \left| \obsa_{lm} \right|^2 \rangle \ = \ \frac{1}{4\pi} \int_{\rm S^2} \d \Omega_{\q_1} \int_{\rm S^2} \d \Omega_{\q_2} \, \langle \obsT (\q_1) \, \obsT (\q_2) \rangle \, P_l(\q_1 \cdot \q_2)  \\
& = &  \sum_{l'=0}^{\lmax} A_{ll'} \, C_{l'} \ + \ C^\text{N}_{l} \, . \label{eq:biasMatDef}
\end{eqnarray}
Here, $\lmax$ is the maximum harmonic multipole,  $C_l^\text{N}$ is the angular power spectrum of noise and the bias matrix,
\begin{equation}
A_{ll'} \ := \ \frac{2l'+1}{16\pi^{2}} \int_{\rm S^2} \d \Omega_{\q_1} \int_{\rm S^2} \d \Omega_{\q_2} P_{l} (\q_1 \cdot \q_2) W_{l'}(\q_1,\q_2) \, ,\label{eq:All'}
\end{equation}
where,
\begin{equation}
W_{l}(\q_1,\q_2) \ := \ \int_{\rm S^2} \d \Omega_{\q'_1} \int_{\rm S^2} \d \Omega_{\q'_2} B(\q_1,\q'_1) \, B(\q_2,\q'_2) \, P_{l}(\q'_1,\q'_2) \, . \label{eq:wl}
\end{equation}
It is interesting to note that the noise-free two-point correlation function of the observed CMB anisotropy sky can be expressed as
\begin{equation}
\langle \obsT(\q) \, \obsT(\q') \rangle \ = \ \sum_{l=0}^{\infty}\frac{(2l+1)}{4\pi}C_{l} \, W_{l}(\q,\q') \, . \label{eq:TT-Cl-relation}
\end{equation}

\subsection{Circular Beams}

It is fairly straightforward to show that if the beam is azimuthally symmetric about the pointing direction, that is, $B(\q,\q') \equiv B(\q \cdot \q')$, so that the beam can be expanded in terms of Legendre polynomials,
\begin{equation}
B(\q,\q') \ \equiv \ B(\q \cdot \q') \ = \ \frac{1}{4\pi} \sum_{l=0}^{\lmax} (2 l \, + \, 1) \, B_l \, P_l(\q \cdot \q') \, ,
\label{eq:Bl}
\end{equation}
the observed angular power spectrum is trivially biased, $A_{ll'} = \delta_{ll'} B_l^2$,
\begin{equation}
\obsC_l \ = \ B_l^2 \, C_l \ + \ C^\text{N}_{l} \, .
\end{equation}
Thus in this case it is easy to get an unbiased estimator using the scalar window function $B_l^2$,
\begin{equation}
\unbsC_l \ = \ B_l^{-2} [\obsC_l \ - \ C^\text{N}_{l}] \, , \label{eq:unbEstSc}
\end{equation}
assuming that the noise power spectrum, $C^\text{N}_{l}$, can be precisely estimated independently from instrument noise characteristics. The co-variance of the unbiased estimators is given by
\begin{equation}
\cov(\unbsC_l, \unbsC_{l'}) \ = \ \frac{2 \delta_{ll'}}{2 l + 1} \left( C_l \ + \ B_l^{-2} C^\text{N}_{l} \right)^2 \, .
\end{equation}
The above equations imply that in case of circular beams, there is no coupling between power spectrum at different multipoles.

Note that, unless otherwise mentioned, we will use a twiddle ( $\widetilde{~}$ )  to denote an observed quantity, a bar ( $\bar{~}$ ) on a quantity to denote an estimator and a superscript ``scalar'' to denote that the estimator is derived using a scalar transfer function.

\subsection{Non-circular Beams}

In general, however, $B(\q,\q')$ can not be expanded in terms of Legendre polynomials, they can be expanded in terms of Spherical Harmonics $Y_{lm}(\q')$ for {\em every pointing direction $\q$}. Making use of Wigner rotation matrices, the spherical harmonic transforms of the beams computed at a specific location could be transformed to other directions~\cite{Souradeep2001}. The final expression for the observed power spectrum can then be expressed in terms of the spherical harmonic transforms of the beam ($b_{lm}$) pointing at the $z$-axis of spherical polar coordinates. Numerical computation then shows that, for a trivial scanning strategy and elliptical Gaussian beams, to the leading order the bias matrix $A_{ll'}$ has a large number of small off diagonal components, which can imply a significantly large difference in power spectrum if estimated assuming a circular beam~\cite{Mitra2004}. Hence, from Eq.~(\ref{eq:biasMatDef}), one should define the true unbiased estimator as
\begin{equation}
\unbC_l \ = \ \sum_{l'} A^{-1}_{ll'} [\obsC_{l'} \ - \ C^\text{N}_{l'}] \, .
\label{eq:unbEst}
\end{equation}

Even though the approximate calculations provide a good estimate of the level of the effect, in order to fully account for the effect one needs the true bias matrix computation involving the actual non-circular beam shape, the exact scan pattern and beyond leading order computation. This is a challenging task and has not yet been accomplished in literature. Efforts are on to compute effective $b_{l2}$ for a complex experiment, incorporating the effect of scanning strategy, in turn enabling one to compute the true bias matrix to a reasonable accuracy using only leading order computation~\cite{PantEtAl}.

\subsection{Scalar effective beam window functions}

Above we described that circular beam window functions can lead to significant systematic error in power spectrum estimation. But, realistic bias matrices for non-circular beams are not available. This situation might seem discouraging ! In practice though there is a middle-ground. One can compute a scalar beam window function through Monte Carlo simulations, incorporating beam asymmetry and scanning strategy. A large number of maps are simulated from a fiducial angular power spectrum, which are convolved with actual non-circular beams. The scalar window function can then be estimated by taking the average ratio of the convolved to unconvolved power spectra. However, a scalar window function corresponds to a circular beam ($b_{lm} \propto \delta_{m0}$), as non-vanishing $b_{lm}$ lead to off-diagonal terms in the bias matrix. So this method essentially replaces the complex effective beams, which can also vary across the sky, by one effective circular beam. The goal of this paper is to study the accuracy and adequacy of this scalar beam window functions in cosmological parameter estimation.

The effective beam window function is defined as the average of the ratio of the convolved ($\obsC_l$) to unconvolved fiducial ($\fidC_l$) simulated maps
\begin{equation}
W_l \ := \ \langle  \obsC_l / \fidC_l \rangle \, . \label{eq:scalarW}
\end{equation}
%
In this paper we simulate a convolved power spectrum by multiplying an unconvolved spectrum ($C_l$) with a given bias matrix
\begin{equation}
\obsC_l \ := \ \sum_{l'=0}^{\lmax}  A_{ll'} \, C_{l'} \, . \label{eq:obsCl}
\end{equation}
This method alleviates the need for performing the Monte Carlo simulations for estimating the scalar transfer function, as neither $A_{ll'}$ nor the fiducial spectrum $C_l$  are random realizations. However, this is possible here because we are restricting ourselves to a case for which the bias matrix is available, which is a primary requirement for this study.
%

Note that, in this paper we will be using two different unconvolved spectra, one is the ``true'' spectrum of the sky, which the analysis aims to recover, the other one is a ``fiducial'' spectrum ($\fidC_{l}$) for estimating the scalar window function via Eq.~(\ref{eq:scalarW}). In both the cases convolution is done through Eq.~(\ref{eq:obsCl}). If one includes the the full bias matrix in the unbiased estimator, as in Eq.~({\ref{eq:unbEst}), the  estimator is {\em truly unbiased}. This is because, combining with Eq.~(\ref{eq:biasMatDef}), one gets
\begin{equation}
\langle  \unbC_l \rangle \ = \ \sum_{l'} A^{-1}_{ll'} [\langle \obsC_{l'} \rangle \ - \ C^\text{N}_{l'}] \ = \ C_l \, .
\end{equation}
However, this need not be the case if one uses a scalar transfer function $W_l$ to define the unbiased estimator, as in Eq.~(\ref{eq:unbEstSc}), because
\begin{equation}
\langle  \unbsC_l \rangle \ = \ W_l^{-1} [\langle \obsC_{l} \rangle \ - \ C^\text{N}_{l}] \ = \ W_l^{-1} \,\sum_{l'=0}^{\lmax}  A_{ll'} \, C_{l'} \, . \label{eq:NewClEstimat}
\end{equation}
Inserting the convention for ``true'' and ``fiducial'' described above one gets
\begin{equation}
\langle  \unbsC_l \rangle \ = \ \fidC_{l} \, \frac{\sum_{l'=0}^{\lmax}  A_{ll'} \, \trueC_{l'}}{\sum_{l'=0}^{\lmax}  A_{ll'} \, \fidC_{l'}} \, .\label{eq:NewClFid}
\end{equation}
It is easy to see from the above equation that if $\fidC_l = \trueC_l$, one would get $\langle \unbsC_l \rangle = \trueC_l$. However, {\em this still does not ensure that the posterior distribution of the parameters can be correctly recovered} by using the best-fit $C_l$ as $\fidC_l$. This is because parameter estimation codes compute likelihood for different sets of $\trueC_l$, obtained by sampling the parameter space, to get the posterior distributions. Hence $\fidC_l$ can match $\trueC_l$ {\em at most} for one realisation of the set of parameters. Moreover, since the aim of an experiment is to estimate the best-fit $C_l$  one can not guess  $\trueC_l$ {\em a priori}. Hence, there is no reason to assume that $\fidC_l = \trueC_l$. Nevertheless, the difference between $\fidC_l$ and $\trueC_l$ need not be very large either, as we do have approximate knowledge of $\trueC_l$ from previous experiments and only the ``nearby'' regions of the parameter space are sampled. In this work, as a realistic test case, we first take $\fidC_l$ to be the WMAP best-fit spectrum and $\trueC_l$ to be the {\Planck} best-fit spectrum and study whether this difference causes significant deviation in cosmological parameter estimation. We then study whether the true posterior distribution can be recovered (not only the best-fit parameter values) by taking $\fidC_l$ and $\trueC_l$ to be the same, the {\Planck} best-fit power spectrum. 


\section{Analysis and Results}
\label{sec:results}

The broad approach we follow here is to generate non-circular beam convolved power spectrum, analyse it including the corresponding bias matrix in the parameter estimation codes and compare the results with the ones obtained using different scalar transfer function.

\subsection{Generation of Convolved Spectrum}

As mentioned above, we generate a convolved singular power spectrum $\obsC_l$ starting from a fiducial spectrum $\fidC_l$ by multiplying the later with the bias matrix $A_{ll'}$ [Eq.~(\ref{eq:obsCl})] corresponding to the chosen non-circular beam. However, the bias matrix $A_{ll'}$ has so far been computed to leading order for very specific cases of elliptical Gaussian beams. We use one such bias matrix for a beam of size and shape crudely similar to that of one of the {\Planck} beams. Note that for this study it is reasonable to assume an approximate bias matrix to be the true bias matrix as the scalar window function is estimated through the same bias matrix.

 We choose the Full Width at Half Maximum (FWHM) of the beam to be $0.1^\circ$ and eccentricity of $\epsilon = 0.7$. In literature often an elliptical Gaussian beam is characterized by ellipticity, the ratio of the semi-major ($a$) to the semi-minor ($b$) axis, while eccentricity is defined as $\epsilon = \sqrt{1-b^2/a^2}$. Hence the ellipticity of the beam used here is $1/\sqrt{1-\epsilon^2} = 1.4$, which is on a slightly higher side compared to the {\Planck} detectors, leading to a more conservative estimate which will become evident later.
 
 We chose a trivial scan pattern that keeps the alignment of the beam fixed with respect to the local meridian over the whole sky. We could not include a realistic scan strategy as the bias matrix for such a case has not been numerically computed  in literature. In fact one of the main aim of this work is to study the need for investing enormous effort in computing the realistic bias matrix. However, this is not a big assumption either for this work. We are studying whether the scalar transfer functions can mimic a typical bias matrix in parameter estimation. Had we incorporated the full scan in the bias matrix, the matrix would be different, but we believe the broad characteristics would remain the same, as can be seen in the bias matrix plots for two different toy scan patterns presented in \cite{Mitra2004}. Hence this study should remain valid for realistic scan patterns.

\subsection{Choice of Scalar Transfer Functions}
\label{sec:scalarW}

We study four different variants of scalar transfer functions ($W_l$) for calculating the estimator of the power spectrum $\unbsC_l := \obsC_l/W_l$ as listed below:
\begin{enumerate}
\item $W_l$ is taken as $B_l^2$, where $B_l$ is the Legendre transform of a circular Gaussian beam, as defined in Eq.~(\ref{eq:Bl}), whose radius is the geometric mean of the semi-major and semi-minor axis of the chosen elliptical Gaussian beam, preserving the total collection solid angle of the beam.
\item $W_l$ is taken as the diagonal components of the of the bias matrix ($A_{ll'}$), i.e. $W_l=A_{ll}$.
\item $W_l$ is computed from a ``fiducial" $C_l$ using Eq.(\ref{eq:scalarW}). We use two different $\fidC_l$
\begin{enumerate}
  \item $\fidC_l$ is taken as the best fit WMAP-9 $C_l$,
  \item $\fidC_l$ is taken as the best fit {\Planck} $C_l$.
\end{enumerate}
\end{enumerate}

\subsection{Modification to the Likelihood Analysis}

We compute the posterior distributions of cosmological parameters from the original $C_l$ and the power spectrum estimators ($\unbC_l$) obtained by using different scalar and tensor transfer functions described above. Parameter estimation is done using the code SCoPE~\cite{Das2014}. We add log-likelihoods from {\tt commander\_v4.1\_lm49.clik}, {\tt lowlike\_v222.clik} and {\tt CAMspec\_v6.2TN\_2013\_02\_26.clik}~\cite{PlanckES-13}. The effect of beam non-circularity is insignificantly small at the low multipoles. Since {\tt commander} and {\tt lowlike} only use $C_l$'s from low multipoles, modifications to $C_l$ with bias matrix or either of the above scalar transfer functions have negligible effect on these likelihoods. The only likelihood that gets affected in this process is that from {\tt CAMspec}. 

{\tt CAMspec} likelihood provides $\chi^2 = (C_l-\unbsC_l)^T [C]^{-1} (C_l-\unbsC_l)$, where $[C]$ is the covariance matrix and $C_l$ is a theoretical spectrum for a given set of parameters. In most situations, to perform parameter estimation for a specific cosmological model, one runs MCMC chains over the parameter space and for each set of parameters $C_l$ and $\chi^2$ are computed. This procedure does not question the validity of the estimator $\unbsC_l$ and its covariance, which is in direct contrast to the situation considered in this work. Here we would like to modify $\unbsC_l$ to capture the effect of using different scalar and tensor beam window functions. However, if we modify $\unbsC_l$ by say $W^{-1}_l\sum_{l'} A_{ll'} \unbsC_{l'}$, the covariance matrix also gets modified to $[W^{-1}_lA_{ll'}][C][W^{-1}_lA_{ll'}]^{T}$. Therefore, for running MCMC with {\tt CAMSpec} likelihood we need to change both $\unbsC_l$ and the covariance matrix and feed those into the {\tt CAMSpec} likelihood code. Rather, we use an alternate approach. For each MCMC realisation we multiply the $C_{l}$'s obtained from CAMB\cite{CAMB} with $W_{l} A_{ll'}^{-1}$, that is, replacing theoretical $C_l$ by $W_{l} \sum_{l'} A_{ll'}^{-1} C_{l'}$ keeping $\unbsC_l$ and $[C]$ fixed. This step provides us the intended $\chi^2$ at a reduced complication. We then estimate the posterior distributions of cosmological parameters from the exact estimator $\unbC_l$ incorporating a bias matrix and the estimators $\unbsC_l$ for different approximations to the beam window described in Section~\ref{sec:scalarW}.

Note that, the above procedure is justified since the aim of this paper is {\em not} a reanalysis of {\Planck} data, which would require a nearly exact treatment of beams and noise, the aim here is to verify if the scalar transfer functions are adequate for representing the distortions described by a bias matrix. So the above procedure in a way assumes that {\Planck} measurements are correctly represented by the supplied beam transfer functions and noise covariance matrix and the default likelihood produces correct values. We distort the theoretical $C_l$ by an inverse bias matrix and check if the effect can be compensated in the likelihood codes by multiplying it with a scalar transfer function. We use {\Planck} likelihood code here only to get realistic noise characteristics and resolution. If one wanted to reanalyse {\Planck} power spectra with bias matrices (assuming that they have somehow been made available for each channel), one would have to modify the observed spectra and the noise covariance matrices separately for each frequency.

\subsection{Results}

We start by plotting the quantity $F_l := \frac{W_l\sum_{l'}A_{ll'}^{-1}C_{l'}^{\rm Planck}}{C_l^{\rm Planck}}$ in figure (\ref{fig:compWin}),  where $C_l^{\rm Planck}$ is the {\Planck} best fit $C_l$ and $W_l$ is estimated in three different (inexact) ways. This plot shows how much error is introduced in the power spectrum estimator for not considering the non-circularity of the beam properly. It can be seen that if we consider the beam to be circular Gaussian then the error in $C_l$ at high multipoles is very large, more than $\sim 3\%$.  However, if we consider the diagonal components of the $A_{ll'}$ as the scalar transfer function then the error involved at high multipoles is lesser, $\sim 1\%$. These levels of error in $C_l$ can affect the cosmological parameters very significantly. On the other hand $W_l$ computed through the forward method from the fiducial $C_l$, as in Eq.(\ref{eq:NewClFid}), performs much better. In the plot we choose the ``fiducial" $C_l$ as $C^{\rm WMAP}_l$. Here the error at high multipoles is less than $0.1\%$. Of course, if $W_l$ was estimated using $C_l^{\rm Planck}$ as the fiducial spectrum, $F_l$ would be $1$ at every $l$. However, as mentioned before, $C_l^{\rm Planck}$ is used as the true spectrum here, which the analysis aims to recover, it can not be guessed a priori. We could also use some other set of parameters to generate the ``true'' spectrum. The {\Planck} best fit parameters are chosen here as they promise to be the closest to reality in the history of CMB measurements.

\begin{figure}[h]
\centering
\includegraphics[width=0.7\textwidth,trim=0.1cm 8.0cm 1.0cm 8.1cm, clip=true]{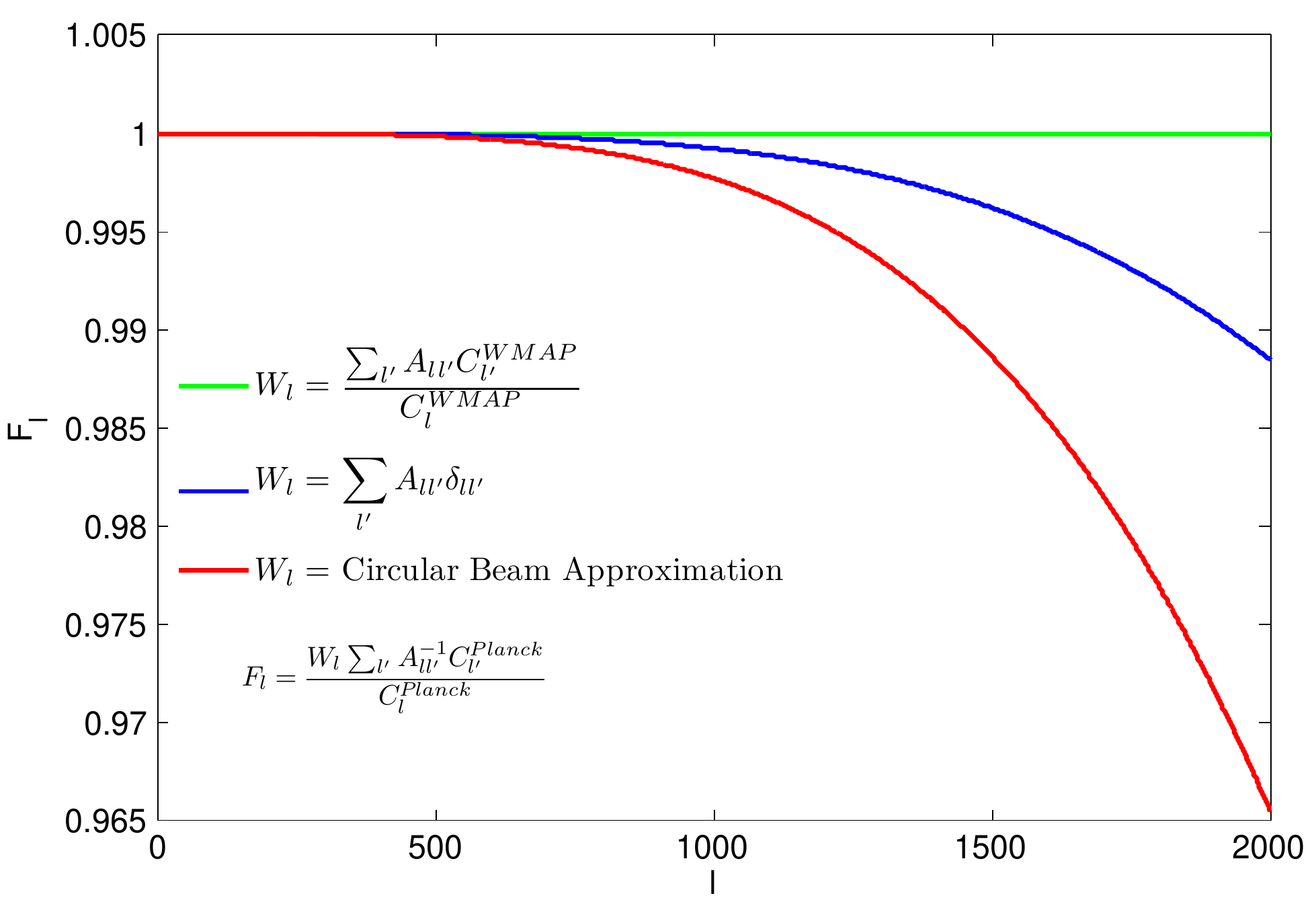}
\caption{\label{fig:compWin} Comparison of different effective window functions. A convolved CMB power spectrum is generated by multiplying the {\Planck} best-fit with a bias matrix for an elliptical Gaussian beam. It is then corrected by a scalar transfer function and its ratio to the original $C_l$ is plotted. The scalar transfer function is obtained in different ways: by assuming a circular Gaussian beam (red), the diagonal of the bias matrix (blue) and using a forward approach with WMAP best-fit as the fiducial $C_l$ (green). If we used {\Planck} best-fit as the fiducial spectrum, the correction would be perfect and we would get $F_l=1$. The good news is that even with WMAP best-fit fiducial $C_l$, $F_l$ is close to unity. The other two scalar transfer functions lead to significant deviation of $F_l$ from unity, implying highly incorrect estimation of the unbiased power spectrum.}
\end{figure}

We have done parameter estimation for the standard six parameter LCDM cosmology, namely \{$\Omega_{c}$,$\Omega_{b}$,$h$,$\tau$,$n_{s}$,$A_{s}$\} using SCoPE~\cite{Das2014}. In figure (\ref{fig:circular}) we show the likelihood contours (in blue) for the six parameters obtained from {\Planck} best fit $C_l$, without introducing any instrumental effect. According to the scheme we followed to introduce the bias matrix, this case is equivalent of accounting for the full bias matrix in the analysis through the exact unbiased estimator [Eq.~(\ref{eq:unbEst})]. Hence, this is the ``correct'' distribution we aim to recover. The contours corresponding to circular Gaussian beam approximation, the first $\unbsC_l$ estimator discussed in Section~\ref{sec:scalarW}, are overlaid (in red). It can be clearly seen that circular Gaussian approximation leads to very different posteriors, in some of the cases the contours get shifted by more than $2\sigma$, implying that the mean values of the parameters can be highly incorrect as well. Using $W_l = A_{ll}$ also leads to similar, but little better, results. The likelihood contours are not shown here, the marginalised distributions are shown in figure~\ref{fig:compare} (in blue).

\begin{figure}[h]
\includegraphics[width=0.95\textwidth]{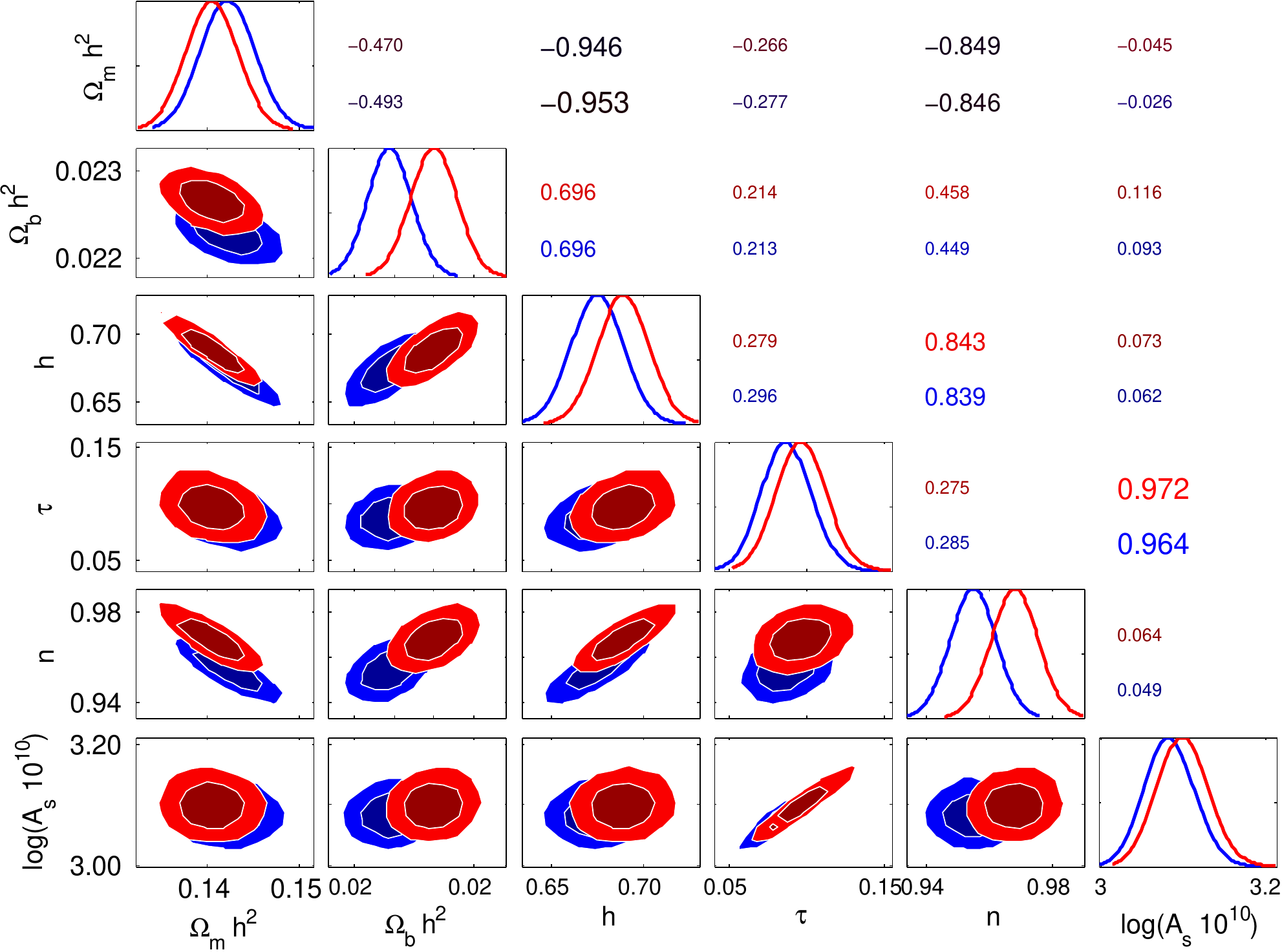}
\caption{\label{fig:circular} We plot the likelihood contours for the six cosmological parameters starting from the correct (unbiased using full bias matrix) pseudo-$C_l$ estimator $\unbC_l$ and a scalar transfer function corrected one ($\unbsC_l$) obtained for circular Gaussian beam. It clearly shows that circular Gaussian approximation for a non-circular beam can lead to very large deviation in the posterior distribution. The correlation coefficients between pairs of parameters are shown in upper triangle of the plot.}
\end{figure}

\begin{figure}[h]
\includegraphics[width=0.95\textwidth]{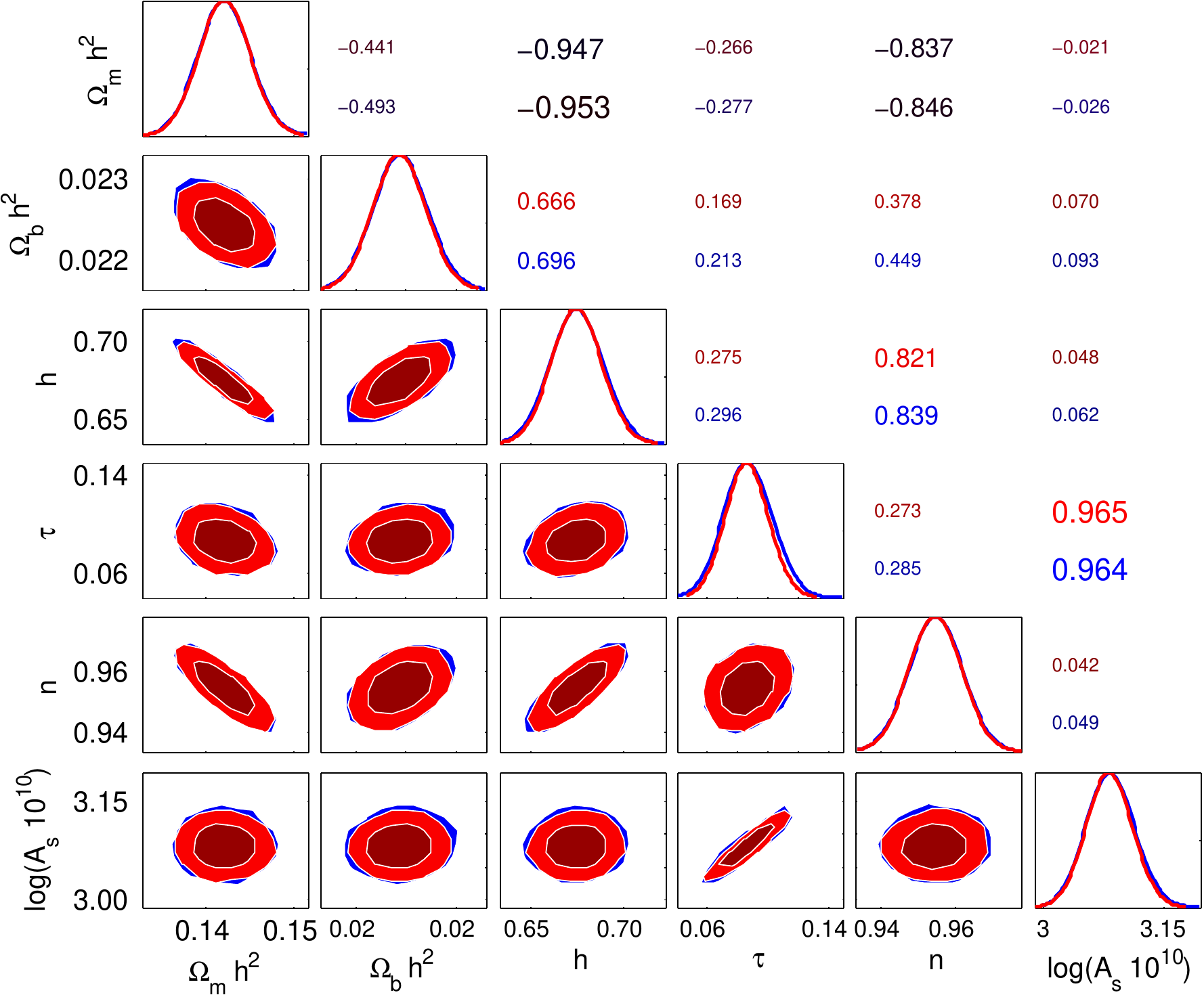}
\caption{\label{fig:planck} We plot the likelihood contours for the six cosmological parameters starting from the correct (bias matrix corrected) pseudo-$C_l$ estimator $\unbC_l$ (in blue) and a scalar transfer function corrected one ($\unbsC_l$), where the transfer function was obtained using {\Planck} best-fit as the fiducial $C_l$ (in red). The correlation coefficients between pairs of parameters are shown in upper triangle of the plot. The plot shows that using a scalar transfer function derived via Monte Carlo simulations provides reasonably close best-fit values, which is clearer in the 1-D marginalized distributions. However, the posterior distributions do show deviations from the actual distributions, which become more significant if WMAP best-fit was used as the fiducial spectrum [see figure~\ref{fig:compare}] instead of {\Planck}, as the later may not be known {\it a priori}. These differences can be reduced by redoing the analysis using transfer functions derived from the best-fit spectrum obtained in the first step, the likelihood contours would then correspond to the red contours in this figure.}
\end{figure}

{\Planck} data analysis uses scalar transfer functions obtained via Monte Carlo simulations starting from a fiducial $C_l$ as mentioned in Section~\ref{sec:scalarW}. We first use the best fit WMAP $C_l$ as the fiducial $C_l$ and use the resulting scalar transfer function corrected power spectrum $\unbsC_l$ for parameter estimation. We find that though the mean parameters are recovered to few percent of sigma accuracy, the distributions are significantly different, which would lead to wrong errorbars. Likelihood contours are not shown for this case, but marginalised distributions are plotted in figure~\ref{fig:compare}. Next we show results with {\Planck} best-fit as the fiducial spectrum. In figure~\ref{fig:planck} the likelihood contours obtained from the full bias matrix corrected $C_l$ are plotted in blue and those from the estimator $\unbsC_l$ are in red. The contours are close for most parameters, but not for all, the 1-D marginalised distributions provide a clearer picture.

\begin{figure}[h]
\centering
\includegraphics[width=0.95\textwidth]{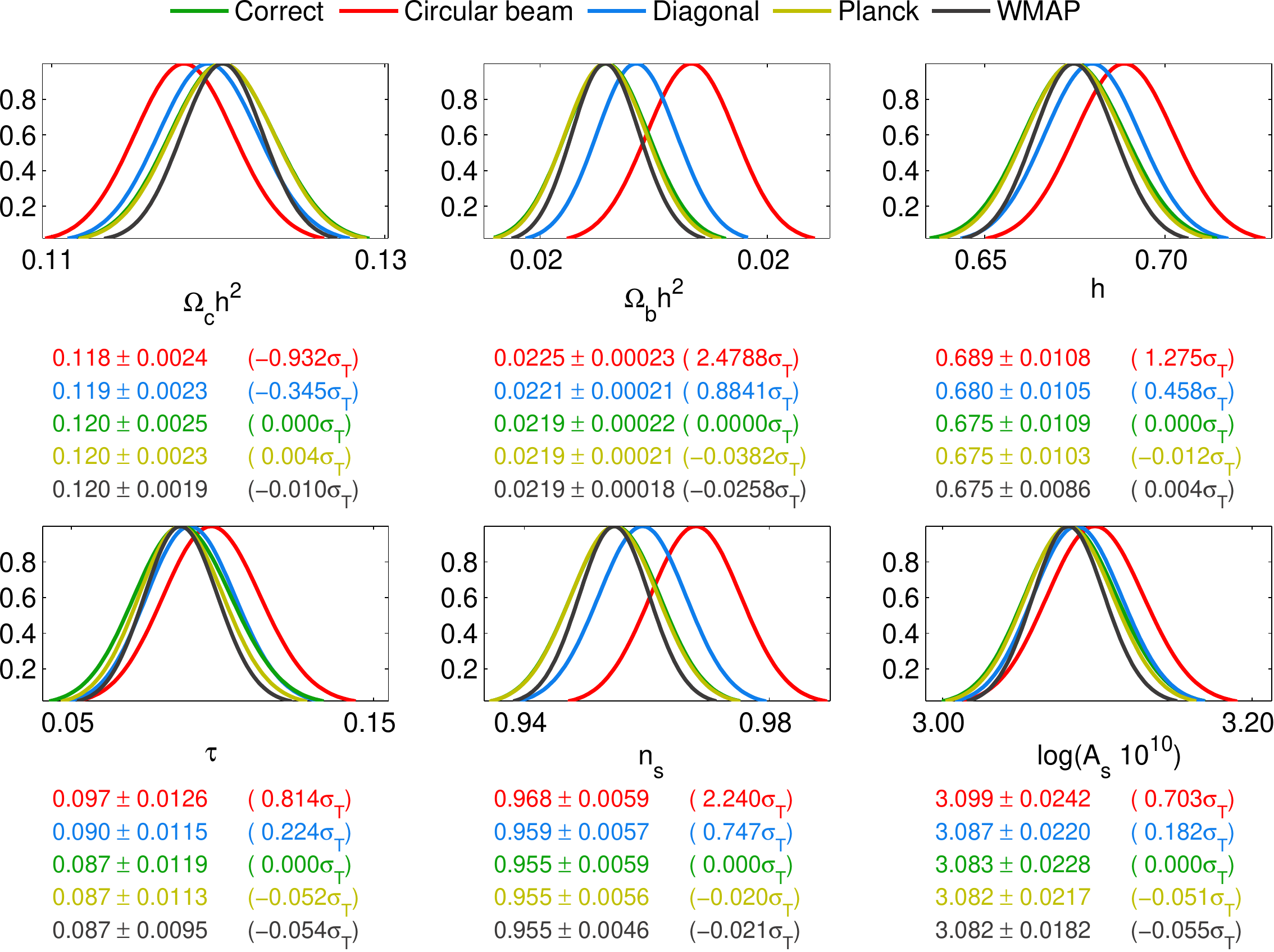}
\caption{\label{fig:compare}One dimensional marginal probability distributions for all the four scalar transfer functions mentioned in Section~\ref{sec:scalarW} are plotted along with the ``correct/true'' distribution that corresponds to the case where the bias matrix is accounted for. We show the average and the standard error below the plots. The deviation of the mean from the correct mean is shown in the bracket in terms of $\sigma_T$, where $\sigma_T$ is the standard deviation of the correct distribution for the parameter in context [that is, in the plot for $n_s$,  $\sigma_T$ is the standard deviation of the distribution for $n_s$ (green)]. Notice that both the yellow and the black marginalised distributions have means shifted only by few percent $\sigma_T$ from the correct analysis, illustrating that the scalar transfer functions derived through Monte Carlo simulations give accurate mean parameters, while the other approximations badly fail. However, the posterior distributions are different, the differences are more significant when the fiducial spectrum used for transfer function estimation is not close to the best fit (compare black and yellow with green).
 }
\end{figure}


In figure~\ref{fig:compare} we overlay the 1-D marginalized probability distributions for the four different cases discussed in section~\ref{sec:results}, along with the distribution from the ``correct'' bias matrix corrected $C_l$ (green). It shows that if the WMAP best fit $C_l$ is used as the fiducial spectrum for deriving the transfer function (black) then also the average values of the distribution are almost the same, but the distributions are narrower, i.e. the standard deviations are smaller than that of the original distribution sometimes by as much as $\sim 20$\%. This implies that the choice of wrong fiducial $C_l$ for obtaining scalar transfer function may lead to wrong distribution of the parameter space, but not significantly changing the mean. When the fiducial spectrum matches the best-fit (yellow), the differences in distributions are reduced and the standard deviations differ by few percent, as in figure~\ref{fig:planck}.

Thus, we can conclude that the method adapted by the {\Planck} team for obtaining the parameters can provide reasonably accurate posteriors, at least if the beams are not too different from elliptical gaussian and the fiducial spectrum is close to the actual best-fit. However, the actual best-fit may not be known {\it a priori}. To address this issue, we suggest running the analysis first with a transfer function derived from a fiducial spectrum constructed from a reasonable initial guess of parameters, obtain the best-fit parameters, recompute the transfer function with the best-fit spectrum and finally get the posterior distribution and update the best-fit by using the new transfer function.


\section{Conclusions and Discussions}
\label{sec:conclusion}

We have estimated cosmological parameters from observed CMB temperature power spectrum by correctly accounting for non-circular beams with the help of a bias matrix and also by using different scalar beam window functions. We show that if the proper beam window function is not used, the power spectrum estimator gets biased and the posterior distribution of parameters becomes significantly different. Namely, assuming a circular Gaussian beam creates serious departure from the true posterior, while using a scalar transfer function estimated through Monte Carlo simulations involving non-circular beams does well in recovering the best fit values to few percent of sigma. However, the posterior distributions show significant differences, leading to incorrect errorbars, if the fiducial spectrum is very different from the actual best-fit. The posterior distributions can be brought closer to actual by introducing an iterative step, where the scalar transfer function is reestimated using the best-fit spectrum obtained in the first step.


We have considered here an elliptical Gaussian beam and a non-rotating scanning strategy, as the bias matrix is available in only in such cases. If the bias matrix was available for a more realistic beam shape and scanning strategy, this work could be immediately repeated for that case. However, we believe that even in such complex situations, the bias matrix would be different but still would not show any dramatically different features. Since we have not used any special characteristics of the bias matrix, this work should still remain valid. Testing that hypothesis of course requires the whole bias matrix to be computed, though this work suggests that, at least for temperature and low-multipole polarization analysis, a full bias matrix may not be necessary for cosmological parameter estimation.

Obtaining the bias matrix for polarization analysis is a even bigger challenge, though efforts are on~\cite{Fidy}. We emphasise that the systematic effects of this kind become important when the errorbars are small even beyond the multipoles corresponding to the beam width. Which happens, for example, in CMB temperature anisotropy measurement with {\Planck}. However, for polarization measurement the errorbars are still large, one may not need such fine corrections for current experiments. 

Here we would also like to mention that for our analysis we only change $C_l^{TT}$, whereas a beam function that affects $C_l^{TT}$ should affect $C_l^{TE}$. However the effect on $C_l^{TE}$ is not significant in the present analysis, considering that the effects of the noncircular beam only affect the high multipoles and presently the polarization and the cross power spectrum measurement is limited low multipoles.

It is well known that masking of foreground contaminated regions in the maps couples the lower multipoles. The full bias matrix for masking with circular beam is routinely used in CMB analysis~\cite{MASTER}. Masking and non-circular beam effects are strong in different multipole ranges, still there can be small coupling between these two effects~\cite{Mitra2009}. However, the effect of this on parameter estimation should be small and may not be significant at the level of precision attainable with current CMB experiments.

The bottomline is that the need for incorporating full bias matrices in parameter estimation may not be crucial with current sensitivities of CMB measurements. Use of scalar transfer functions derived through Monte Carlo simulations, similar to what was done in {\Planck} analysis, provides sufficient accuracy in best-fit parameter estimation and can also provide somewhat accurate posterior distributions if handled carefully. However, efforts should be invested in estimating the full bias matrices in these complex situations for putting more precise bounds on the cosmological parameters. Moreover, polarised bias matrices must be computed, to the leading order to start with, at least for performing a study like this one to make a statement about the adequacy of scalar transfer functions in high resolution polarisation analysis in {\Planck} and post-{\Planck} era.

\acknowledgments{
We would like to thank Krzysztof G\'orski  and Tarun Souradeep for useful discussions. We have used the HPC facility at IUCAA for the required computation. SD acknowledge Council for Science and Industrial Research (CSIR), India, for the financial support as Senior Research Fellows. SM acknowledges the support of the Science and Engineering Research Board (SERB), India through the Fast Track grant SR/FTP/PS-030/2012.}

\bibliographystyle{JHEP}
\bibliography{beamPar}

\end{document}